\documentclass{agujournal}

\newcommand{\blue}[1]{{{{\color{black} #1}}}}

\usepackage{hyperref}
\usepackage{amssymb,amsfonts,amsmath}
\usepackage{multirow}
\usepackage{tabularx}
\usepackage{color}
\usepackage{graphicx}
\usepackage{amsthm}
\usepackage[toc,page]{appendix}

\usepackage[letterpaper, lmargin=0.3in, rmargin=0.8in]{geometry}

\journalname{Journal of Geophysical Research}

\begin{document}

%
%


\title{Multivariate quadrature for representing cloud condensation nuclei activity of aerosol populations}

%
%





\authors{Laura Fierce\affil{1,2} and Robert L. McGraw\affil{1}}
 \affiliation{1}{Climate and Environmental Sciences Department, Brookhaven National Laboratory}
 \affiliation{2}{Visiting Scientist Programs, University Corporation for Atmospheric Research}



\correspondingauthor{Laura Fierce}{lfierce@bnl.gov}




\begin{keypoints}
\item Multivariate quadrature resolves key features of multi-component aerosol populations
\item Constrained maximum entropy distribution are used to reconstruct critical supersaturation distributions 
\item The combined technique reproduces CCN spectrum from particle-resolved model with high accuracy
\end{keypoints}

%
%


\begin{abstract}
Sparse representations of atmospheric aerosols are needed for efficient regional- and global-scale chemical transport models. Here we introduce a new framework for representing aerosol distributions, based on the quadrature method of moments. Given a set of moment constraints, we show how linear programming, combined with an entropy-inspired cost function, can be used to construct optimized quadrature representations of aerosol distributions. The sparse representations derived from this approach accurately reproduce cloud condensation nuclei (CCN) activity for realistically complex distributions simulated by a particle-resolved model. Additionally, the linear programming techniques described in this study can be used to bound key aerosol properties, such as the number concentration of CCN. Unlike the commonly used sparse representations, such as modal and sectional schemes, the maximum-entropy approach described here is not constrained to pre-determined size bins or assumed distribution shapes. This study is a first step toward a particle-based aerosol scheme that will track multivariate aerosol distributions with sufficient computational efficiency for large-scale simulations.
\end{abstract}

%
%


\section{Introduction}
Atmospheric aerosols remain a large source of uncertainty in prediction of changes and variability in the climate system \citep{myhre2001historical, kaufman2002satellite, carslaw2013large, stocker2013ipcc}. Particle characteristics relevant for climate, such as optical cross sections and the critical supersaturation at which particles serve as cloud condensation nuclei (CCN), depend on particle size, shape, and chemical composition \citep[e.g.][]{jacobson2001strong,chung2011effect,cubison2008influence,ervens2010ccn}. Observations show tremendous variability in particle physical and chemical properties \citep[e.g.][]{krieger2012exploring, zhang2014variation}, but complex distributions of multicomponent particles are not easily represented in large-scale models. In this study, we seek a sparse representation of aerosol populations that resolves enough information about particle size and composition distributions to adequately represent climate-relevant properties, while resolving as little information as is necessary in order to minimize computational costs.

Many numerical models have been developed to simulate the aerosol evolution \citep[e.g.][]{wexler1994modelling,mcgraw1997description,jacobson1997development,penner1998climate,binkowski2003models,stier2005aerosol,bauer2008matrix,riemer2009simulating}. The simplest models track particle mass only, without any resolution of particle size or composition \citep[e.g.][]{haywood1997general,myhre1998estimation,penner1998climate,cooke1999construction,lesins2002study}. At the other extreme, the Particle Monte Carlo Model (PartMC) \citep{riemer2009simulating} coupled to the Model for Simulating Aerosol Interactions and Chemistry (MOSAIC) \citep{zaveri2008model} tracks the evolution of thousands of individual particles, but is computationally expensive. Many large-scale models now include aerosol microphysical schemes that simulate the evolution of the particle size distribution, but resolve limited information about particle composition. Sectional models track the evolution of particles in different size bins, typically assuming that particles of the same size have identical composition \citep[e.g.][]{wexler1994modelling,jacobson1997development}. Sectional models have also been extended to resolve multivariate distributions with respect to particle composition, such as particle hygroscopicity parameter or BC mass fraction, in addition to particle size \citep{oshima2009aging,matsui2013development,zhu2015size}, or to track separate aerosol populations \citep{kleeman20013d,jacobson2002analysis}. Because sectional schemes simulate the evolution of a discretized representation of the full particle size distribution or, in the multivariate case, the particle size-composition distribution, they require tracking a large number of bins to resolve aerosol populations.

Moment-based models are a class of methods for simulating the evolution of probability density functions in which integral quantities over the distribution, such as radial moments, are tracked, rather than evolving the distribution itself \citep{hulburt1964some,mcgraw1997description}. In this way, moment-based models represent the aerosol using an intermediate level of detail between the simple mass-based aerosol schemes and the more detailed particle-resolved and sectional representations. To date, moment-based aerosol representations have been implemented in global models as modal or monodisperse schemes that track only two low-order radial moments for each population \citep[e.g.][]{stier2005aerosol,bauer2008matrix}, typically aerosol number and volume or mass. Monodisperse aerosol models follow a similar approach \citep{pirjola2003monodisperse}. Modal models begin by assuming a specific distribution shape, typically lognormal, track two moments per mode, and assume a prescribed geometric standard deviation. Particles within a mode are further assumed to contain identical mass fractions of the constituent aerosol species. However, comparison between modal and sectional aerosol schemes show that modal models can yield inaccurate representation of the particle size distribution \citep{mann2012intercomparison}. Further, inadequate resolution of variability in composition within each mode leads to errors in prediction of climate-relevant properties \citep{fierce2016black,fierce2016toward}.


\blue{The present study introduces a new framework for constructing aerosol representations in moment-based models,} including modal models. Focusing on particle CCN activation properties, traditionally a difficult case for moment-based representations \citep{wright2001description,wright2002retrieval}, we show how efficient quadrature representations can be constructed from moments of aerosol distributions bivariate with respect to dry diameter and hygroscopicity parameter. By applying a transformation of variables we construct sparse quadrature point distributions of the aerosol in terms of the critical saturation for cloud droplet activation. Continuous distributions are obtained off line using the maximum entropy spectral representation constrained by the same moment/quadrature parameters that defined the sparse distribution. We show that this approach yields accurate prediction of CCN activity using only a limited number of moments.

\section{Constructing sparse and continuous representations from distribution moments}\label{sec:methods}
Here we describe a new method for utilizing moments and moment-based, multivariate quadrature approximation to obtain sparse representations of benchmark PartMC-generated aerosol populations. Full size distributions, derived in terms of the sparse quadrature representation, are obtained using the principle of maximum entropy. Putting these two steps together we demonstrate that multivariate quadratures can be used to accurately compute CCN activities for the benchmark population.



\subsection{Computing CCN activity for benchmark populations}\label{sec:method_benchmark}
The particle-resolved model PartMC-MOSAIC was used to generate realistically complex particle populations (see Appendix~\ref{sec:appendix_benchmark}), which were used to benchmark the new method. \blue{PartMC-MOSAIC tracks the masses of 20 aerosol components, including water, in thousands of computational particles, corresponding to a stochastic representation of the 20-dimensional distribution that describes aerosol mixing state.} The procedure for computing CCN activity from a particle-resolved population is shown in Figure~\ref{fig:flowchart}a--\ref{fig:flowchart}c. The particle-resolved simulations reveal complex variation in particle size and composition, as illustrated by the number density distribution with respect to two coordinates: dry diameter $D_{\text{dry}}$, which is the volume-equivalent diameter for all species except water, and effective hygroscopicity parameter $\kappa$, which is the volume-weighted hygroscopicity parameter for each constituent aerosol species \citep{petters2007single}. \blue{The bivariate distribution $n(D_{\text{dry}},\kappa)$ is the reduced form of the 20-dimensional distribution with respect to aerosol components that is most relevant for CCN activation, but the procedure described in this study can be extended to different projections of the particle size-composition distribution for aerosol dynamics simulations and calculation of aerosol optical properties.} The bivariate number distribution $n(D_{\text{dry}},\kappa)$ for the benchmark population is shown in Figure~\ref{fig:flowchart}a, where $n(D_{\text{dry}},\kappa)$ is normalized by the total particle number concentration. Because the particle-resolved model resolves composition at the particle level, it is uniquely suited for benchmarking approximate aerosol representations, including the quadrature-based approach discussed here. 

For the particle-resolved population, the normalized number distribution with respect to critical supersaturation, $n(s_{\text{c}})$, is shown in Figure~\ref{fig:flowchart}b. The critical supersaturation at which each simulated particle becomes CCN-active, $s_{\text{c}}$, is computed as a function of $D_{\text{dry}}$ and $ \kappa$ (see Appendix~\ref{sec:appendix_sc}). The overall number fraction of particles able to serve as CCN at each supersaturation threshold $N_{\text{CCN}}(s)/N$, where $N$ is the total particle number concentration, is then computed as the cumulative distribution of $n(s_{\text{c}})$, shown for this benchmark population in Figure~\ref{fig:flowchart}c.

\subsection{Constructing quadratures from moments of $n(D_{\text{dry}},\kappa)$}\label{sec:method_quadPoints}
Figure~\ref{fig:flowchart}d shows the location of quadrature abscissas for a specific quadrature approximation for the bivariate distribution shown in Figure~\ref{fig:flowchart}a, derived from a specific set of moment constraints $\mu_k$. A generalized moment $\mu_k$ of the distribution $n(D_{\text{dry}},\kappa)$ is an integral over $n(D_{\text{dry}},\kappa)$ with respect to some kernel function of the coordinates, $\phi_k(D_{\text{dry}},\kappa)$:
\begin{equation}\label{eqn:constraints}
\mu_k=\int_0^{\infty}\int_0^{\infty}n(D_{\text{dry}},\kappa)\phi_k(D_{\text{dry}},\kappa)dD_{\text{dry}}d\kappa.
\end{equation}
For example, for $\phi_k(D_{\text{dry}},\kappa)=D_{\text{dry}}^{m}$, $\mu_k$ is the \blue{$m^{\text{th}}$} \blue{power moment with respect to diameter} and, therefore, independent of $\kappa$. Here, we computed constraints in terms of modified moments \citep{press1990numerical}, which are linear combinations of the power moments (see Appendix~\ref{sec:appendix_moments}). Continuity equations, expressed in terms of moments, can be solved exactly in closed form only in very special cases, but approximate closure can be achieved by representing integrals over the number distribution using numerical quadrature. \blue{Remarkably}, to be accurate, quadrature-based models require tracking only a small number of quadrature abscissas and weights\blue{;} the distribution itself is not required. The quadrature approximation of $n(D_{\text{dry}},\kappa)$ takes the form:
\begin{equation}\label{eqn:quadrature}
\mu_k\approx\sum_{i=1}^Nw_i\phi_k(D_{\text{dry},i},\kappa_i),
\end{equation}
such that $\mu_k$, expressed in terms of the bivariate \blue{abscissa coordinates}, $D_{\text{dry},i}$ and $\kappa_i$, and weights, $w_i$, is identical to the result of Equation~\ref{eqn:constraints} for $k=1,...,N_{\text{q}}$. \blue{In this way, for each} $\mu_k$, the summation over $N_{\text{q}}$ quadrature points in Equation~\ref{eqn:quadrature} is exact. Algorithms have been developed to construct quadrature approximations that satisfy power moments and modified moments of univariate distributions \citep{gordon1968error,press1990numerical}, and these approximations yield on the order of $N_{\text{q}}/2$ quadrature points for $N_{\text{q}}$ moment constraints. This paper introduces a new technique for generating a quadrature representation from sets of generalized constraints, which provides a distributed set of $N_{\text{q}}$ quadrature points for $N_{\text{q}}$ generalized moments and is also suitable for constructing quadrature for multivariate distributions.

We seek a quadrature approximation that can be derived from a set of known constraints $\mu_k$. Ideally, the quadrature abscissas $x_i$ will be distributed across the variable space in order to resolve key features of the probability distribution. We found that a distributed set of abscissas, \{$D_{\text{dry},i}$, $\kappa_i$\}, and weights, $w_i$, can be constructed through a linear program that maximizes an entropy-inspired cost function (see Appendices~\ref{sec:appendix_entropy}~and~\ref{sec:appendix_linearprogram}).

\subsection{Reconstructing $n(s_{\text{c}})$ from quadrature}\label{sec:method_reconstructDist}
\begin{figure}
\begin{center}
\includegraphics[width = 3in,trim=0.9in 0 0.67in 0,clip=true]{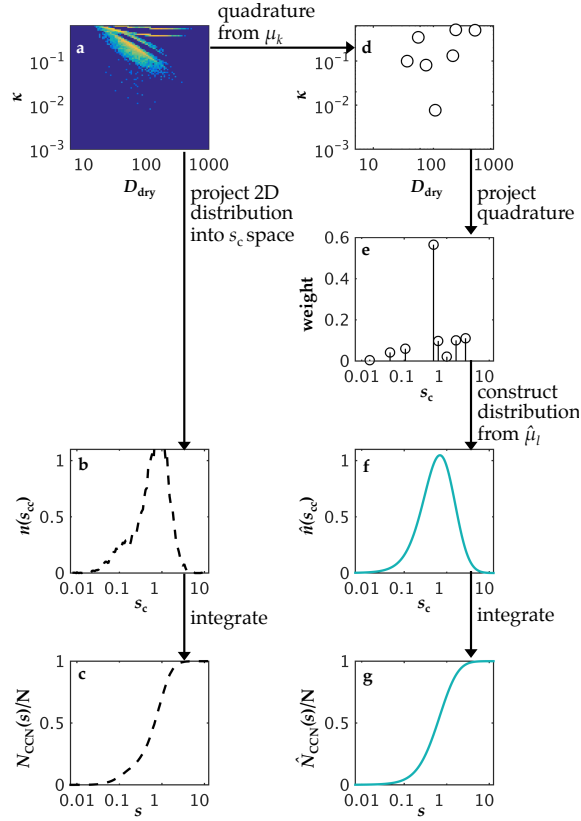}
\caption[flowchart]{\label{fig:flowchart} Procedure for computing CCN spectrum from particle-resolved model (a--c) and from quadrature-based representation (d--g): (a) bivariate number distribution with respect to $D_{\text{dry}}$ and $\kappa$ from particle-resolved model, (b) number distribution with respect to $s_{\text{c}}$ from particle-resolved model, (c) $N_{\text{CCN}}(s)/N$ from particle-resolved model, (d) location of abscissas in $D_{\text{dry}}$-$\kappa$ space for quadrature, (e) abscissas transformed into $s_{\text{c}}$ and weights for quadrature, (f) maximum entropy reconstruction of number density with respect to $s_{\text{c}}$ from quadrature, and (g) $N_{\text{CCN}}(s)/N$ for quadrature.}
\end{center}
\end{figure}

For aerosol representations that track only quadrature points (Figure~\ref{fig:flowchart}b), a procedure is required if one seeks a corresponding continuous distribution, and often this can be offline. We propose to transform each quadrature point from the bivariate abscissas, \{$D_{\text{dry},i}$, $\kappa_i$\}, into univariate abscissas, $s_{\text{c},i}$, for each quadrature point $i=1,...,N_{\text{q}}$ \blue{and preserving the weights}, as shown in Figure~\ref{fig:flowchart}.c. The moments of $n(s_{\text{c}})$, $\mu_l$ can then be estimated using the transformed quadrature: 
\begin{equation}\label{eqn:sc_quadrature}
\mu_l=\sum_{i=1}^Nw_i\phi_l(s_{\text{c},i}).
\end{equation}
The number of moments in $s_{\text{c}}$ space, $N_{s}$, need not be the same as the number of moments in $D_{\text{dry}}$-$\kappa$ space, $N_{\text{q}}$. 


The desired continuous distribution $n(s_{\text{c}})$, shown in Figure~\ref{fig:flowchart}.f, is constructed by finding the distribution having maximum entropy, given the approximated moments in $s_{\text{c}}$ space, $\hat{\mu}_l$ for $l=1,...,N_{\text{s}}$ (see Appendix~\ref{sec:appendix_entropy}). \blue{We use $\hat{\mu}_l$ to indicate the moments in $s_{\text{c}}$ space approximated from the quadrature and $\mu_l$ to indicate exact moments computed from the benchmark population, which would not be known in a quadrature-based simulation.} Similarly, we denote the reconstructed distribution as $\hat{n}(s_{\text{c}})$ to differentiate between the benchmark distribution $n(s_{\text{c}})$. If only the moments of a distribution are known, the distribution that best represents \blue{the state of knowledge afforded by the moment constraints is the distribution having maximal entropy \citep{jaynes1957information,jaynes1957information2}}. Distributions of maximum entropy can be constructed for any set of moment constraints and are not limited to univariate aerosol distributions, \blue{but here we confine ourselves to the univariate case}. \blue{In Section~\ref{sec:optimalMoments}, we show that the accuracy of the reconstructed distribution depends strongly on the set of moments $\mu_l$ chosen to reconstruct the distribution and the accuracy with which the moments approximated from the quadrature $\hat{\mu}_l$ represent the moments taken directly from the particle-resolved distribution $\mu_l$.}

\subsection{Evaluating error in $N_{\text{CCN}}(s)/N$}
The aim of this study is to find a moment-based representation that accurately reproduces CCN activation properties of benchmark aerosol populations from the particle-resolved model PartMC-MOSAIC. To this end, we apply a genetic algorithm (see Appendix~\ref{sec:appendix_geneticalgorithm}) to find optimal sets of bivariate moments to be constrained in constructing the bivariate quadrature. We define $N_{\text{CCN}}(s)$ and $\hat{N}_{\text{CCN}}(s)$ as the number concentration of CCN from the particle-resolved model and from a quadrature representation, respectively, and $N$ is the overall number concentration of particles, which is constrained to be identical in both cases. The error $\varepsilon$ is quantified as the mean squared error:
\begin{equation}
\varepsilon = \frac{1}{N}\int_0^{\infty}\big(\hat{N}_{\text{CCN}}(s)-N_{\text{CCN}}(s)\big)^2ds.
\end{equation}
The following sections describe methods for identifying \blue{optimized} sets of moment constraints for constructing the quadrature in $D_{\text{dry}}$-$\kappa$ space $\mu_k$ and \blue{optimized} sets of moments for reconstructing continuous distributions in $s_{\text{c}}$ space. We define \blue{optimized} moment sets as those yielding the best representation of the CCN activation spectrum, indicated by small mean squared error $\varepsilon$.

\section{Moment constraints for aerosol simulations}\label{sec:optimalMoments}
Although the proposed maximum-entropy approaches for constructing quadrature approximations and continuous distributions can yield accurate representations of the benchmark populations, here we show that this accuracy depends strongly on the selection of moment constraints. \blue{The moments with respect to $s_{\text{c}}$ computed from the bivariate quadrature, $\hat{\mu}_l$ are used to construct the continuous distribution $\hat{n}(s_{\text{c}})$} (see Section~\ref{sec:method_reconstructDist}); Section~\ref{sec:optimalMoments_sc} introduces the necessary and sufficient set of constraints $\mu_l$ to accurately reconstruct $n(s_{\text{c}})$. The accuracy of the continuous distribution $\hat{n}(s_{\text{c}})$ depends on the degree to which the projected quadrature points, which are originally constructed in $D_{\text{dry}}$-$\kappa$ space (see Section~\ref{sec:method_quadPoints}), approximate the true moments of $n(s_{\text{c}})$ for the benchmark distribution; Section~\ref{sec:optimalMoments_diaKap} describes a procedure for selecting moment constraints of $n(D_{\text{dry}},\kappa)$ required to construct \blue{accurate} quadrature approximations of $n(D_{\text{dry}},\kappa)$ and $\hat{n}(s_{\text{c}})$.

\subsection{Moments of $n(s_{\text{c}})$ for reconstructing distributions}\label{sec:optimalMoments_sc}
In this section, we identify the set of moments, $\mu_l$ for $l=1,..,N_{\text{s}}$, that \blue{yield highest accuracy for the reconstructed distribution} $\hat{n}(s_{\text{c}})$. We show that the accuracy of the reconstruction depends on the degree to which the quadrature is able to reproduce the moments taken directly from the particle-resolved distribution $n(s_{\text{c}})$. 

\begin{figure}
\begin{center}
\includegraphics[width = 2.5in]{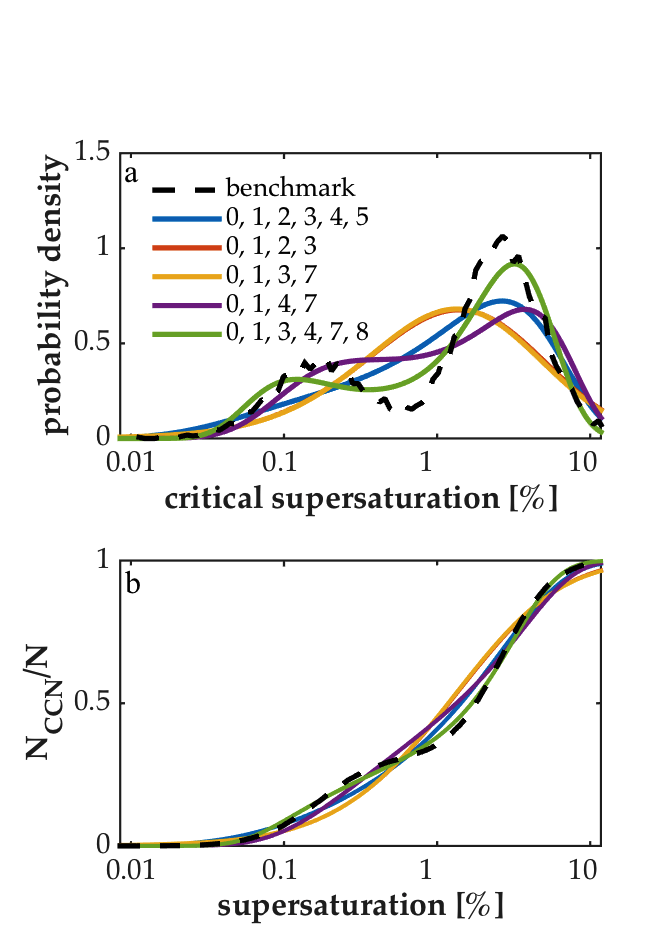}
\caption[Sc_moments]{\label{fig:Sc_moments} Comparison between benchmark (black dashed lines) and reconstructions (colored lines) of (a) number density with respect to $s_{\text{c}}$ and (b) $N_{\text{CCN}}(s)/N$ \blue{for different combinations of moments of $n(s_{\text{c}})$, computed directly from the benchmark population, reveals that only~six moments $\mu_l$ are needed, but these moments must be carefully selected.}}
\end{center}
\end{figure}

Figure~\ref{fig:Sc_moments}a shows distributions constructed from the quadrature using selected combinations of moments $\mu_l$. We found that $n(s_{\text{c}})$ can be estimated with high accuracy using the six modified moments $\mu_l$, corresponding to Legendre polynomials of power $l=0,1,3,4,7,8$, \blue{computed with respect to $\ln{s}_{\text{c}}$} (see Appendix~\ref{sec:appendix_moments}). Provided these six moments can be computed with high accuracy from projections of the quadrature approximation of $n(D_{\text{dry}},\kappa)$, the moment-based representation reproduces the CCN spectrum $N_{\text{CCN}}(s)/N$ of the benchmark populations with high accuracy, as shown through comparison between green and black lines in Figure~\ref{fig:Sc_moments}b.


\subsection{Moments of $n(D_{\text{dry}},\kappa$) for constructing bivariate quadrature}\label{sec:optimalMoments_diaKap} 
A good bivariate quadrature representation of the aerosol must reproduce key moments of $n(s_{\text{c}})$ on projection, which are used to construct the CCN activation spectrum $N_{\text{CCN}}(s)/N$. Selecting the optimal six moments from ten possible combinations, as was done for selecting key moments of $n(s_{\text{c}})$, requires testing a total of 210 combinations. On the other hand, selecting optimal combinations of bivariate moments requires testing unfeasibly large sets of combinations. For example, here we identify combinations of bivariate Legendre moments of power $n=-3,..,8$ (Equation \ref{eqn:legendre_DdryKap}) for each of the two variables, $D_{\text{dry}}$ and $\kappa$, yielding 144 possible bivariate moments. Selecting the eight optimal moments from the 144 candidates would require performing the full procedure for \mbox{$3.8\times10^{11}$} combinations. Instead, we applied a genetic algorithm to find suitable sets of bivariate moments $\mu_k$, $k=1,..,N_{\text{q}}$, \blue{defined by Equations~\ref{eqn:constraints}~and~\ref{eqn:quadrature}}, that should be tracked in moment-based models for accurate representation of CCN properties. 

\begin{figure}
\begin{center}
\includegraphics[width = 2.5in]{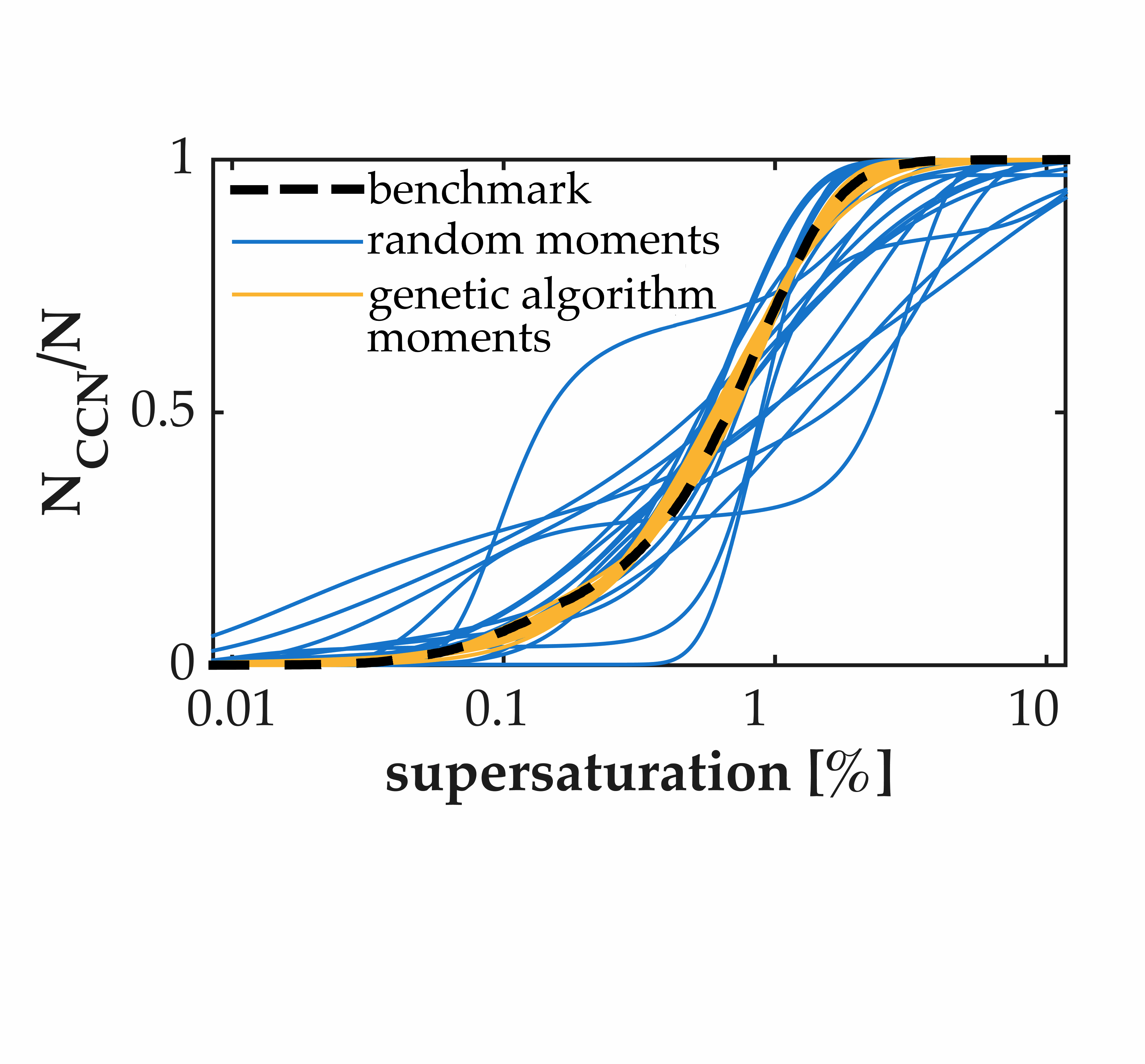}
\caption[DiaKap_moments]{\label{fig:DiaKap_moments} Comparison between benchmark (black dashed line), reconstructions from randomly selected moments (blue lines), and reconstructions from moments chosen with the genetic algorithm (orange lines) show that care must be taken in selection of of bivarriate moments constraints for construction of the quadrature.}
\end{center}
\end{figure}

For a single population, the CCN spectra computed from the quadrature, $\hat{N}_{\text{CCN}}(s)/N$, is compared with the particle-resolved CCN spectrum, $N_{\text{CCN}}(s)/N$ in Figure~\ref{fig:DiaKap_moments}, using 20~randomly selected combinations (blue lines) of $N_{\text{q}}=8$ moments and for 20~combinations that were selected using the genetic algorithm (orange lines). Although the proposed procedure yields high accuracy in approximated CCN spectra $\hat{N}_{\text{CCN}}(s)/N$ for those moments selected by the genetic algorithm, the large errors in $\hat{N}_{\text{CCN}}(s)/N$ for randomly chosen moments illustrate that the accuracy of the procedure depends strongly on the choice of moment constraints used in assignment of the bivariate quadrature points. 

\section{Sparse approximation for efficient and accurate aerosol representation}
\begin{figure}
\begin{center}
\includegraphics[width = 5.5in]{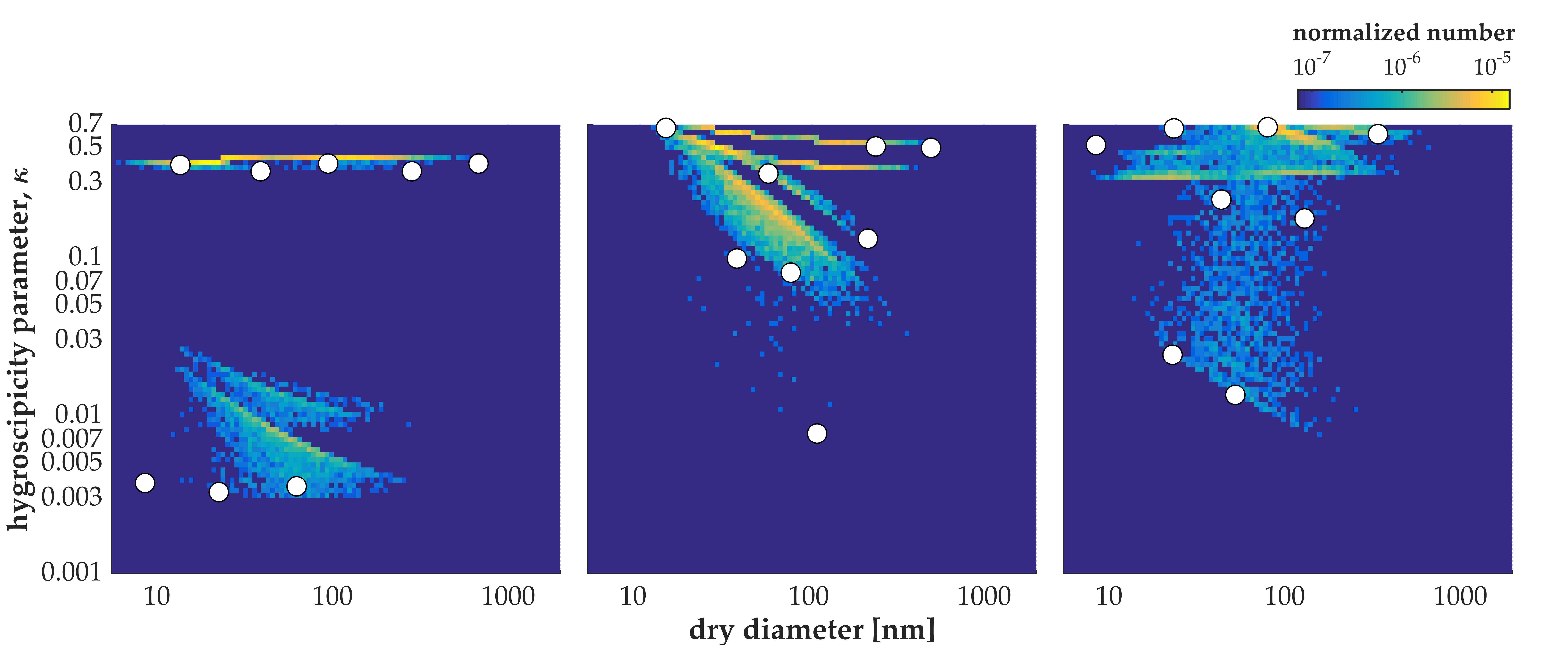}
\caption[DiaKap_moments]{\label{fig:DiaKap_quad} Bivariate distributions from particle-resolved model (surface plots) and location of abscissas from the quadrature (white dots) for the three populations, which were sampled at (a) 7:00~am, after 1~hour of simulation, (b) 12:00~pm, after 6~hours of simulation, and (c) 6:00~am the following day, after 24~hours of simulation. The corresponding weights are not shown here but are shown by the white dots in the one-dimensional projection plots of Figure~\ref{fig:Sc_reconstruct}. In all cases, the maximum-entropy-inspired technique yields a distributed set of abscissas.}
\end{center}
\end{figure}

For the selected moments of $n(D_{\text{dry}},\kappa)$, $\mu_k$, and the selected univariate moments for estimating $n(s_{\text{c}})$, $\mu_l$, the quadrature-based representation accurately reproduces CCN activation for the three bivariate benchmark populations shown in Figure~\ref{fig:DiaKap_quad}. The \blue{location of quadrature abscissas} associated with a set of $N_{\text{q}}=8$ moments (white dots) are superimposed on benchmark distributions (surface plots). The quadrature was constructed using the approach outlined in Section~\ref{sec:method_quadPoints} for moments of $n(D_{\text{dry}},\kappa)$, where the specific combinations of moment constraints $\mu_k$ for $k=1,...,N_{\text{q}}$ were identified using the genetic algorithm (Section~\ref{sec:optimalMoments_diaKap} and Appendix~\ref{sec:appendix_geneticalgorithm}). \blue{Optimized bivariate moments identified with the genetic algorithm are listed in Table~\ref{tab:optimized_moments} of Appendix~\ref{sec:appendix_geneticalgorithm}.} The quadrature represents key features of each benchmark distribution using only moments.

\begin{figure}
\begin{center}
\includegraphics[width = 5.5in]{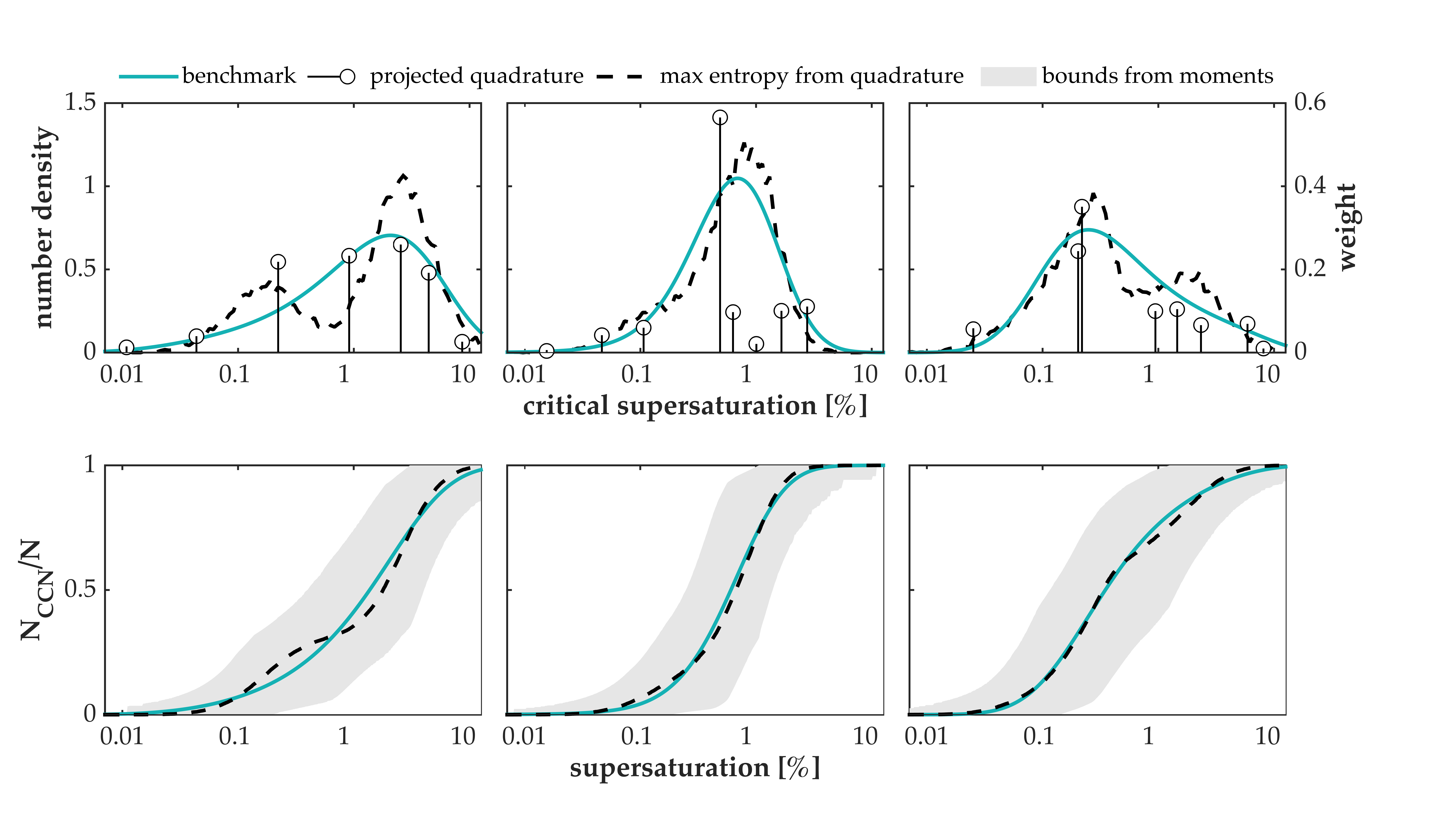}
\caption[DiaKap_moments]{\label{fig:Sc_reconstruct} (a--c) The number distribution with respect to $s_{\text{c}}$ for the benchmark (dashed curves) are compared with reconstructions from the quadrature (blue curves), where the projection of the underlying quadrature abscissas and weights, \blue{indicated by the location and magnitude of the vertical stems.} (d--f) The bivariate moments used to construct the quadrature do not effectively constrain $N_{\text{CCN}}(s)/N$, where bounds on possible solutions are shown by the shading, but the continuous distributions reconstructed from the quadrature (blue curves) accurately represent the benchmark (black dashed curves). Results are shown for the three particle-resolved populations from Figure~\ref{fig:DiaKap_quad}, which were sampled at (a,d) 7:00~am, after 1~hour of simulation, (b,e) 12:00~pm, after 6~hours of simulation, and (c,f) 6:00~am the following day, after 24~hours of simulation.}
\end{center}
\end{figure}

The critical supersaturation $s_{\text{c},i}$ is computed for each quadrature abscissa $\{D_{\text{dry},i},\kappa_i\}$, shown along with corresponding weights by the stems in Figures~\ref{fig:Sc_reconstruct}a--\ref{fig:Sc_reconstruct}c. These quadrature points are used to compute the projected moments in $s_{\text{c}}$ space (Section~\ref{sec:optimalMoments_sc}), and the full distribution $\hat{n}_{\text{c}})$ is then reconstructed from the estimated moments (Section~\ref{sec:method_reconstructDist}). The reconstructed distributions (blue lines in Figures~\ref{fig:Sc_reconstruct}a--\ref{fig:Sc_reconstruct}c) represent key features of the benchmark distributions (black lines). Although the distributions $\hat{n}(s_{\text{c}})$ are not reconstructed exactly, the CCN activation spectra, $\hat{N}_{\text{CCN}}/N$, that is computed by integrating over $\hat{n}(s_{\text{c}})$ reproduces the benchmark spectrum $N_{\text{CCN}}(s)/N$ with high accuracy (Figures~\ref{fig:Sc_reconstruct}d--\ref{fig:Sc_reconstruct}f).

The shaded region of Figures~\ref{fig:Sc_reconstruct}d--\ref{fig:Sc_reconstruct}f show the bounds on possible solutions for each CCN spectrum, given the modified moments $\mu_k$ that were used to construct the quadrature approximations in Figure~\ref{fig:DiaKap_quad}; all possible distributions that satisfy the moment constraints $\mu_k$ lie within these extreme bounds. The large bounds in the shaded region indicate that constraints on $n(D_{\text{dry}},\kappa)$ do not, on their own, sufficiently constrain CCN activation properties. \blue{On the other hand, adding the maximum-entropy-inspired approach described here accurately represents CCN activation spectra across various particle-resolved distributions.}


\section{Discussion and Conclusions}
This manuscript introduces a new procedure for inverting between moments, quadratures, and full distributions, which can be used to advance quadrature-based simulations of aerosol dynamics. By applying a linear program, with a maximum-entropy-inspired cost function, we generated sets of distributed quadrature points that are an \blue{ideal model representation of atmospheric aerosols}, where a genetic algorithm was used to identify optimal combinations of modified bivariate moments to be used as constraints in the linear program. We found that CCN activation spectra can be computed with high accuracy by reconstructing univariate distributions with respect to $s_{\text{c}}$, where moment constraints of $n(s_{\text{c}})$ were determined using a variable transform from bivariate quadrature points $\{D_{\text{dry},i},\kappa_i\}$ to univariate points $s_{\text{c},i}$. The combined procedure yields high accuracy in CCN spectra in comparison with particle-resolved benchmark populations. \blue{Although the present manuscript illustrates the efficacy of the proposed procedure for estimating the CCN activation spectra, which is notoriously difficult for moment-based aerosol representations \citep{wright2001description,wright2002retrieval}, the approach is not limited to the variable spaces that we have chosen to analyze here. In a future study, we will demonstrate how the procedure can be applied to advance quadrature-based aerosol simulations and for computing other quantities of interest, such as aerosol optical properties.} 

In this study, the traditional quadrature method of moments has been augmented with a maximum-entropy-inspired linear program for quadrature generation and with maximum-entropy spectral analysis to obtain continuous representations of consistent with specified moment constrains. We show that the full particle-resolved data set of over 10,000 particles can be reduced to a sparse set of just eight weighted particles to achieve accurate recovery of CCN activation properties. We show that the quadrature method of moments, combined with maximum entropy methods, can be used to overcome the limitations of traditional moment methods.

\begin{appendices}\label{sec:appendix}
\section{Benchmark Populations from Particle-Resolved Model}\label{sec:appendix_benchmark}
PartMC-MOSAIC simulates the evolution of trace gases and aerosol particles in a well-mixed air parcel. The model tracks the mass of $j=1,...,A$ constituent aerosol species in each particle $i=1,...,N_{\text{p}}$, where $A$ is the total number of aerosol species, including water, and $N_{\text{p}}$ is the total number of particles in the population. In the present implementation $A=20$ and $N_{\text{p}}$ is on the order of $10^5$ particles. The mass of each component in each particle is given by $m_{j,i}$, and the mass composition of each particle is represented by the vector $\hat{m}_i=[m_{1,i},...m_{j,i},...,m_{A,i}]$. The model simulates the aerosol evolution due to condensation and evaporation of semi-volatile gases, coagulation between particles, dilution of background air, and particle emissions. A full description of the model is described in \citep{riemer2009simulating}, and the bivariate distributions with respect to $D_{\text{dry}}$ and $\kappa$ are introduced in \citet{fierce2013cloud}.

\section{Critical supersaturation for CCN activation}\label{sec:appendix_sc}
The critical supersaturation ($s_{\text{c},i}$) was computed for each particle $i$ using the $\kappa$-K\"{o}hler model, which depends on a particle's overall dry volume, its effective hygroscopicity parameter $\kappa_i$, and environmental properties. The equilibrium saturation ratio ($S_i$) over an aqueous droplet is computed through the $\kappa$-K\"{o}hler model \citep{petters2007single} as:
 \begin{equation}
\label{eqn:Kohler}
S_i(D_i)=\frac{D_i^3-D_{\text{dry},i}^3}{D_i^3-D_{\text{dry},i}^3(1-\kappa_i)}\exp\left(\frac{4\sigma_{\text{w}}M_{\text{w}}}{RT\rho_{\text{w}}D_i}\right),
\end{equation}
where the wet particle volume $D_i$ is the particle wet diameter, $T$ is the ambient temperature, $R$ is the universal gas constant, $M_{\text{w}}$ is the molecular weight of water, and $\sigma_{\text{w}}$ is the surface tension of the air-water interface. The effective hygroscopicity parameter $\kappa_i$ is the volume-weighted average of the hygroscopicity parameter $\kappa$ of the particle's constituent aerosol species:
\begin{equation}\label{eqn: kappa}
\kappa_i = \frac{\sum_k^{A-1}{v_{k,i}\kappa_k}}{\sum_k^{A-1}{v_{k,i}}}.
\end{equation}
Values for $\kappa$ for each species are given in \citep{fierce2013cloud}.

The critical wet diameter is the diameter at which $S_{i}(D_i)$ is maximal, and the critical saturation ratio ($S_{\text{c},i}$) is the saturation ratio corresponding to this critical wet diameter. The critical supersaturation $s_{\text{c},i}$ is then given by $s_{\text{c},i}=(S_{\text{c},i}-1)\times100$. The number concentration of CCN at each environmental supersaturation $s$ is computed as the number concentration of particles that will activate into at that $s$, that is the total number concentration of particles per volume having a critical supersaturation $s_{\text{c},i}\le{s}$.

\section{Modified Moments from Legendre Polynomials}\label{sec:appendix_moments}
This study uses modified moments based on Legendre polynomials, which are better conditioned than traditional geometric moments. The $l^{\text{th}}$ modified moment over the univariate critical supersaturation distribution $n(s_{\text{c}})$ is defined as the integral over the $l^{\text{th}}$ Legendre polynomial $\phi_l$ of $\ln{s}_{\text{c}}$:
\begin{equation}\label{eqn:legendre_sc}
\phi_{l}(s_{\text{c}}) = \frac{1}{2^ll!}\bigg[\frac{d^l}{d(\ln{s}_{\text{c}})^l}(\ln^2{s}_{\text{c}}-1)^l\bigg].
\end{equation}
Legendre polynomials are defined for $l\ge0$. We define modified moments of $n(s_{\text{c}})$ for \mbox{$l<0$} as integrals over the $l^{\text{th}}$ Legendre polynomials for $\ln{s}_{\text{c}}^{-1}$. \blue{For example, the Legendre polynomials of order 3 and $-3$ are given by \mbox{$\phi_3({s}_{\text{c}})=\frac{1}{2}(5\ln^3{s}_{\text{c}}-3\ln{s}_{\text{c}})$} and \mbox{$\phi_{-3}({s}_{\text{c}})=\frac{1}{2}(5\ln^{-3}{s}_{\text{c}}-3\ln^{-1}{s}_{\text{c}})$}, respectively.}

For the bivariate distributions $n(D_{\text{dry}},\kappa)$, each modified moment $\mu_k$, is computed as the integral over the kernel function $\phi_k(D_{\text{dry}},\kappa)$. We define the modified kernel function $\phi_k$ as the product of the $m^{\text{th}}$ Legendre polynomial of $\ln{D}_{\text{dry}}$ and the $n^{\text{th}}$ Legendre polynomial of $\ln{\kappa}$:
\begin{equation}\label{eqn:legendre_DdryKap}
\phi_{k}(D_{\text{dry}},\kappa)=\bigg(\frac{1}{2^mm!}\bigg[\frac{d^m}{d(\ln{D}_{\text{dry}})^m}(\ln^2{D}_{\text{dry}}-1)^m\bigg]\bigg)\bigg(\frac{1}{2^nn!}\bigg[\frac{d^n}{d(\ln{\kappa})^n}(\ln^2{\kappa}-1)^n\bigg]\bigg).
\end{equation}

\section{Distributional Entropy}\label{sec:appendix_entropy}
The entropy $H_s$ of the reconstructed univariate distribution $\hat{n}(s_{\text{c}})$, which is normalized by total particle number concentration, is given by:
\begin{equation}
H_{s} = \int_0^{\infty}\hat{n}(s_{\text{c}})\ln{\hat{n}}(s_{\text{c}})d{\ln{s}}_{\text{c}}.
\end{equation}
\citet{jaynes1957information,jaynes1957information2} showed that if $n(s_{\text{c}})$ is the density distribution having maximum entropy, given a set of constraints $\mu_l$ for $l=1,...,N_{\text{s}}$, $n(s_{\text{c}})$ will take the form:
\begin{equation}\label{eqn:maxentropy_sc_distribution}
\hat{n}(s_{\text{c}}) = \exp\bigg(-\sum_{l=1}^{N_{\text{s}}}\lambda_l\phi_l(s_{\text{c}})\bigg),
\end{equation}
where $\lambda_l$ is the Lagrange multiplier of the entropy function $H_{\text{s}}$ for $\mu_l$. 

Similarly, the entropy $H_{\text{q}}$ for the bivariate number density distribution $n(D_{\text{dry}},\kappa)$ is given by:
\begin{equation}\label{eqn:bivariate_entropy}
H_q = \int_{-\infty}^{\infty}\int_{-\infty}^{\infty}n(D_{\text{dry}},\kappa)\ln n(D_{\text{dry}},\kappa)d\ln{D}_{\text{dry}}d\ln\kappa.
\end{equation}

\section{Linear Program to Compute Quadrature Approximation and CCN bounds}\label{sec:appendix_linearprogram}
A linear program was used to construct the optimized quadrature, with abscissa locations shown in Figure~\ref{fig:DiaKap_quad}, and to bound $N_{\text{CCN}}(s)/N$, shown by shading in Figure~\ref{fig:Sc_reconstruct}. The linear program maximizes some cost function $c(D_{\text{dry},i},\kappa_i)$ subject to specified constraints $\mu_k$ (Equation~\ref{eqn:constraints}) and the requirement that $w_i\ge0$:
\begin{align}
 & \text{maximize} &\quad-& \sum_{i=1}^{N_{\text{grid}}}w_ic(D_{\text{dry},i},\kappa_i) \\
 & \text{subject to} &\quad& \sum_{i=1}^{N_{\text{grid}}}w_i\phi_k(D_{\text{dry},i},\kappa_i)=\mu_k,\\
                 &&& w_i\ge0, \quad i = 1,...,N_{\text{grid}}.
\end{align}

The application of linear programming to construct numerical quadrature was introduced in \citet{mcgraw2013sparse}, but we now extend this approach to find optimized sets of quadrature points by maximizing an entropy-inspired cost function:
\begin{equation}
c(D_{\text{dry},i},\kappa_i)=-\sum_k\lambda_k\phi_k(D_{\text{dry},i},\kappa_i).
\end{equation}

The bounds on $N_{\text{CCN}}(s)/N$ are computed using two linear programs at each supersaturation threshold $s$, \blue{one for the upper bound and one for the lower bound}. The maximum value for $N_{\text{CCN}}(s)/N$ for some specified $s$ is computed by maximizing the number of particles with $s_{\text{c},i}\le{s}$, which is computed from the linear program with the following cost function:
\begin{equation}
c(D_{\text{dry},i},\kappa_i)=s_{\text{c}}(D_{\text{dry},i},\kappa_i)\le{s}.
\end{equation}
Similarly, the minimum $N_{\text{CCN}}/N$ is computed using a cost function that maximizes the number of particles with $s_{\text{c},i}\ge{s}$:
\begin{equation}
c(D_{\text{dry},i},\kappa_i)=s_{\text{c}}(D_{\text{dry},i},\kappa_i)\ge{s}.
\end{equation}

The linear program is applied on a grid $x_i$ for $i=1,...,N$, where $N$ is much larger than the number of moment constraints. Using the dual simplex algorithm \citep{lemke1954dual}, the linear program yields only a subset of non-zero weights $w_i$, where the number of non-zero values for $w_i$ is equal to the number of moment constraints. It is these non-zero weights $w_i$ and associated abscissas $x_i$ that comprise the sparse set of quadrature points. 

\section{Genetic Algorithm to Optimize Constraints}\label{sec:appendix_geneticalgorithm}
We applied a genetic algorithm to find optimal sets of moments for constructing the bivariate quadrature. Genetic algorithms are a class of optimization methods inspired by natural selection, first introduced by \citet{holland1975adaptation}. Implementation of a genetic algorithm requires a genetic encoding of possible solutions, in this case as a series of 0's and 1's, and fitness function to be optimized, in this case minimization of error in $N_{\text{CCN}}(s)/N$ across distinct populations. The algorithm is initiated with a population of candidate solutions. In this study, the initial set of candidate populations was chosen using Latin hypercube sampling. The optimal order of Legendre polynomials for computing the modified moments with respect to $\ln{D}_{\text{dry}}$ and $\ln\kappa$ (Equation~\ref{eqn:legendre_DdryKap}) identified with the genetic algorithm are outlined in Table~\ref{tab:optimized_moments}.


\begin{table}
\centering
\caption[table]{\label{tab:optimized_moments} Kernel function $\phi_k$ used to generate optimized bivariate moments $\mu_k=\sum_i^{N_{\text{q}}}\phi_k$, which were identified using the genetic algorithm.}
\begin{tabular}{l}
\hline 
$\phi_1=\frac{1}{8}\big(35\ln^4D_{\text{dry}}-30\ln^2D_{\text{dry}}+3\big)$ \\ 
$\phi_2=\ln^{-1}\kappa$ \\ 
$\phi_3=\frac{1}{2}\big(3\ln^{-2}\kappa-1\big) $ \\
$\phi_4=\frac{1}{2}\big(3\ln^{2}D_{\text{dry}}-1\big)\ln\kappa $ \\
$\phi_5=\ln{D_{\text{dry}}}\frac{1}{2}\big(3\ln^{-2}\kappa-1\big) $ \\
$\phi_6=\frac{1}{2}\big(3\ln^{-2}D_{\text{dry}}-1\big)\frac{1}{2}\big(3\ln^{-2}\kappa-1\big) $ \\
$\phi_7=\frac{1}{2}\big(3\ln^{-2}D_{\text{dry}}-1\big)\frac{1}{8}\big(35\ln^4\kappa-30\ln^2\kappa+3\big) $ \\
$\phi_8=\frac{1}{8}(63\ln^5D_{\text{dry}}-70\ln^3D_{\text{dry}}+15\ln{D}_{\text{dry}}) \frac{1}{8}\big(63\ln^5\kappa-70\ln^3\kappa+15\kappa\big)$ \\
\hline 
\end{tabular}
\end{table}
\end{appendices}

\section*{Acknowledgements}
LMF is supported by the University Corporation for Atmospheric Research under a NOAA Climate \& Global Global Change Postdoctoral Fellowship. RLM and LMF acknowledge support by the Atmospheric Systems Research Program of the US Department of Energy. The model (PartMC Version 2.1.6) and input files for the particle-resolved model simulations are available at \url{http://lagrange.mechse.illinois.edu/partmc/}. All other methods supporting the conclusions are described in Section~\ref{sec:methods} and in the Appendices.

 \newcommand{\noop}[1]{}

\listofchanges

\end{document}